\documentclass[prd, aps, superscriptaddress, preprintnumbers, twocolumn, floatfix, nofootinbib]{revtex4}
\pdfoutput=1

\usepackage{amsmath}
\usepackage{tikz}
\usepackage{mathdots}
\usepackage{yhmath}
\usepackage{cancel}
\usepackage{color}
\usepackage{siunitx}
\usepackage{array}
\usepackage{multirow}
\usepackage{amssymb}
\usepackage{gensymb}
\usepackage{tabularx}
\usepackage{booktabs}
\usetikzlibrary{fadings}
\usepackage{amsfonts}
\usepackage{amsmath}
\usepackage{amssymb}
\usepackage{bm}
\usepackage{dcolumn}
\usepackage{graphicx}   
\usepackage[latin1]{inputenc}
\usepackage{latexsym}
\usepackage{rotating}
\usepackage{hyperref}
\usepackage{graphicx}
\usepackage{xcolor, cancel}
\usepackage[normalem]{ulem}
\newcommand\be{\begin{equation}}
\newcommand\ba{\begin{eqnarray}}
\newcommand\ee{\end{equation}}
\newcommand\ea{\end{eqnarray}}

\begin{document}
\preprint{YITP-SB-19-29}

\title {Trans-Planckian Censorship and Inflationary Cosmology}

\author{Alek Bedroya}
\email{abedroya@g.harvard.edu}
\affiliation{Jefferson Physical Laboratory, Harvard University, Cambridge, MA, 02138, USA}

\author{Robert Brandenberger}
\email{rhb@physics.mcgill.ca}
\affiliation{Department of Physics, McGill University, Montr\'{e}al, QC, H3A 2T8, Canada}

\author{Marilena Loverde}
\email{marilena.loverde@stonybrook.edu}
\affiliation{C.N. Yang Institute for Theoretical Physics, Department of Physics \& Astronomy, Stony Brook University, Stony Brook, NY, 11794, USA}

\author{Cumrun Vafa}
\email{vafa@g.harvard.edu}
\affiliation{Jefferson Physical Laboratory, Harvard University, Cambridge, MA, 02138, USA}

\date{\today}

\begin{abstract}

We study the implications of the recently proposed {\it Trans-Planckian Censorship Conjecture} (TCC) for early universe cosmology and in particular inflationary cosmology. The TCC leads to the conclusion that if we want inflationary cosmology to provide a successful scenario for cosmological structure formation, the energy scale of inflation has to be lower than $10^9$ GeV. Demanding the correct amplitude of the cosmological perturbations then forces the generalized slow-roll parameter $\epsilon$ of the model to be very small ($<10^{-31}$). This leads to the prediction of a negligible amplitude of primordial gravitational waves. For slow-roll inflation models, it also leads to severe fine tuning of initial conditions.

\end{abstract}

\maketitle

\section{Introduction} 
\label{sec:intro}

Cosmological observations provide detailed information about our Universe on the largest observable scales. Cosmic microwave background (CMB) measurements \cite{Gorski:1996cf,Bennett:2012zja, Aghanim:2019ame}, for instance, demonstrate that fluctuations in the matter and energy persist on cosmological scales. There is no causal explanation for the origin of these fluctuations in Standard Big Bang cosmology. Scenarios of early universe cosmology such as the Inflationary Universe \cite{Guth} provide a causal mechanism to generate these fluctuations. A key aspect of both inflationary cosmology and of other scenarios that provide an explanation for the origin of structure in the universe (see e.g. \cite{RHBrev} for a comparative review) is the existence of a phase in the early universe during which the Hubble horizon $H^{-1}(t)$, where $H(t)$ is the Hubble expansion rate as a function of time $t$, shrinks in comoving coordinates. The Hubble horizon provides the limiting length above which causal physics that is local in time cannot create fluctuations. In both inflationary cosmology and in the proposed alternatives, comoving length scales which are probed in current cosmological experiments were inside the Hubble horizon at early times.  It is postulated that fluctuations in both matter \cite{ChibMukh} and gravitational waves \cite{Starob} originate as quantum vacuum perturbations that exit the Hubble radius during the early phase, are squeezed and classicalize, and then re-enter the Hubble radius at late times to produce the CMB anisotropies and matter density perturbations that we observe today. 

In \cite{Jerome} (see also \cite{Niemeyer}) it was realized that if the inflationary phase lasts somewhat longer than the minimal period, then the length scales we observe today originate from modes that are smaller than the Planck length during inflation. This was called the {\it trans-Planckian problem} for cosmological fluctuations (see also \cite{Greene}).  This problem was viewed not so much as an issue with a particular model, but more as a question of how to treat trans-Planckian modes in such a situation.  It has been conjectured \cite{Bedroya:2019snp}, however, that this trans-Planckian problem can never arise in a consistent theory of quantum gravity and that all the models which would lead to such issues are inconsistent and belong to the Swampland.  This is called the {\it Trans-Planckian Censorship Conjecture} (TCC).

According to the TCC no length scales which exit the Hubble horizon could ever have had a wavelength smaller than the Planck length.  In Standard Big Bang cosmology no modes ever exit the Hubble horizon, and the TCC has no implications (indeed the TCC is automatically satisfied for all models with a $w\geq {-1/3}$). However, in all early universe scenarios which can provide an explanation for the origin of structure, modes exit the Hubble horizon in an early phase. If $a_i$ is the value of the cosmological scale factor at the beginning of the new early universe phase, and $a_f$ is the value at the time of the transition from the early phase to the phase of Standard Big Bang expansion, the TCC reads
\be \label{TCC}
\frac{a_f}{a_i} \, < \,  \frac{M_{pl}}{H_f} \, ,
\ee
where $H_f$ is the radius of the Hubble horizon at the final time $t_f$ and $M_{pl}$ is the reduced Planck mass. 

In \cite{Bedroya:2019snp}, the relationship between the TCC and other {\it swampland conjectures} \cite{swamp} which have recently attracted a lot of attention (see e.g. \cite{swamprevs} for reviews) was discussed. Here, we focus on the consequences of the TCC for inflationary cosmology.

It is clear from the form of (\ref{TCC}) that the TCC will have strong implications for inflationary cosmology. In the case of de Sitter expansion, $a_f$ is exponentially larger than $a_i$, and hence (\ref{TCC}) strictly limits the time duration of any inflationary phase. The implications for some alternative early universe scenarios are weaker. For example, in {\it String Gas Cosmology} \cite{BV}, the early phase is postulated to be quasi-static. Hence, the condition (\ref{TCC}) is satisfied: no modes which were larger than the Hubble scale at the beginning of the Standard Cosmology phase ever had a wavelength smaller than the Planck length. The same is true for the various bouncing scenarios (e.g. the  {\it matter bounce} \cite{FB}, and the {\it Pre-Big-Bang} \cite{PBB} and {Ekpyrotic} \cite{Ekp} scenarios)\footnote{Of course it is possible that these models have issues related to other Swampland conditions.}, where the initial phase is one of contraction. This is true as long as the energy scale at the bounce point is smaller than the Planck scale. In the following we will study the consequences of the TCC for inflationary cosmology.

The outline of this paper is as follows: In the following section we discuss general constraints imposed by the TCC on the energy scale of inflation and the resulting consequences for the amplitude of gravitational waves. These conclusions do not depend on what drives inflation, only that it occurred. In Section 3, we then specialize to slow-roll inflation models, and show that consistency with the TCC leads requires fine-tuning of the initial conditions. 

We will work in the context of homogeneous and isotropic cosmologies with $4$ space-time dimensions. For simplicity, we assume spatially flatness so that the metric can be written in the form
\be \label{FRW}
ds^2 \, = \, dt^2 - a(t)^2 d{\bf x}^2 \, ,
\ee
where ${\bf x}$ are the spatial comoving coordinates and $a(t)$ is the scale factor (which can be normalized such that $a(t_0) = 1$, where $t_0$ is the present time). The Hubble expansion rate is
\be
H(t) \, \equiv \, \frac{{\dot{a}}}{a} \, ,
\ee
and its inverse is the Hubble radius. As is well known (see \cite{MFB} for an in-depth review of the theory of cosmological fluctuations and \cite{RHBfluctrev} for an overview), quantum-mechanical fluctuations oscillate on sub-Hubble scales, whereas they freeze out and become squeezed when the wavelength is larger than the Hubble radius. During a phase of accelerated expansion, the proper wavelengths of fluctuations initially smaller than the Hubble scale can be stretched to super-Hubble scales. This transition from sub-Hubble to super-Hubble is referred to as horizon-crossing. The TCC prohibits horizon-crossing for modes with initial wavelengths smaller than the Planck length.

We will be considering models of inflation in which a canonically normalized scalar field $\phi$ with potential energy $V(\phi)$ constitutes the matter field driving the accelerated expansion of space. We will be using
units in which the speed of sound, Boltzmann's constant and $\hbar$ are set to 1.

\section{Implications of the TCC for the Energy Scale of Inflation}

In this section we work in the approximation that the Hubble expansion rate during the period
of inflation is constant. In order for inflation to provide a solution to the structure formation problem of Standard Big Bang cosmology, the current comoving Hubble radius must originate inside
the Hubble radius at the beginning of the period of inflation (see Fig. 1). This condition reads
\be \label{success}
\frac{1}{H}\cdot e^{N_+} \cdot \frac{a_{R}}{a_{end}}  \cdot \frac{T_R\  g_*(T_R)^{1/3}}{T_0 \ g_*(T_0)^{1/3}}\simeq \, \frac{1}{H_0} ,
\ee
where $H$ is the Hubble scale during inflation, $a_{\rm{end}}$ and $a_R$ are the values of the scale factor at the
end of inflation and when reheating is completed, and $g_{*}$ indicates
the number of spin degrees of freedom in the thermal bath.
Here $N_{+}$ is the number of e-foldings accrued during the inflation after the CMB-scale modes exit the horizon, $T_0$ is the temperature of the CMB at the present
time, and $T_R$ is the corresponding temperature after reheating.   Equation \eqref{success} can be summarized as follows. We start with a Hubble horizon scale which at the time of the inflation is $1/H$, by the end of inflation it is magnified by $e^{N_+}$, by reheating it has grown again by $a_R/a_{\rm end}$, and between reheating and the present day it grows by the ratio of the $T_R/T_0$  (corrected by the number of degrees of freedom). To solve the horizon problem, this scale should then be larger than the Hubble scale of the current universe $1/H_0$.
To obtain the order of magnitude
of the constraints, we consider
rapid reheating and take the reheating temperature to be
given by the potential energy at the end of inflation, and
hence set $a_R \sim a_{\rm{end}}$.
For simplicity we also assume that the ratio of $g_*$'s is 1.

Under the assumption that the
period of reheating lasts less than one Hubble time $T_R$ is given by the potential energy $V$
during inflation via $T_R \approx V^{1/4}$. Using the Friedmann equation, the Hubble scale $1/H_0$ is given by the
current energy density $\rho_0$ via
\be
\frac{1}{H_0} \, = \, \sqrt{3}\rho_0^{-1/2} M_{pl} \, .
\ee
In turn, $\rho_0$ can be re-expressed in terms of the temperature $T_0$:
\be
\rho_0 \, \approx \, T_0^4 \frac{T_{eq}}{T_0}{1\over \Omega_m}  \,
\ee
where $\Omega_m$ is the fraction of energy density in matter and
 $T_{eq}$ is the temperature at the time of equal matter and radiation, and we
have used the fact that the matter energy density today is larger than the radiation energy
density $T_0^4$ by the factor $T_{eq} / T_0$. Using $H = V^{1/2}/(\sqrt{3} M_{pl})$,
the condition (\ref{success}) then becomes
\be \label{success2}
e^{N_+} \, \simeq \, \frac{V^{1/4}}{(T_0 T_{eq})^{1/2}}\sqrt{\Omega_m} \sim \frac{V^{1/4}}{(T_0 T_{eq})^{1/2}}  \, .
\ee

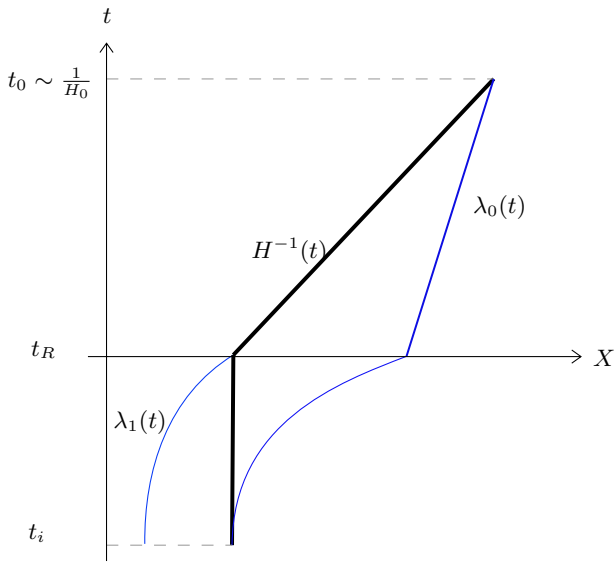
\begin{figure}[t] 
\begin{center}

\begin{tikzpicture}[x=0.5pt,y=0.5pt,yscale=-1,xscale=1]

\draw [color={rgb, 255:red, 8; green, 59; blue, 245 }  ,draw opacity=1 ]   (230.5,423) .. controls (230,367) and (247.5,311) .. (297.5,280) ;

\draw [line width=1.5]    (494.5,71) -- (297.5,280) ;

\draw [color={rgb, 255:red, 0; green, 0; blue, 0 }  ,draw opacity=1 ][line width=1.5]    (297.5,280) -- (296.5,424) ;

\draw [color={rgb, 255:red, 155; green, 155; blue, 155 }  ,draw opacity=1 ] [dash pattern={on 4.5pt off 4.5pt}]  (201.5,71) -- (494.5,71) ;

\draw [color={rgb, 255:red, 15; green, 17; blue, 243 }  ,draw opacity=1 ]   (296.5,424) .. controls (297.5,331) and (376.5,301) .. (428.5,281) ;

\draw [color={rgb, 255:red, 13; green, 14; blue, 226 }  ,draw opacity=1 ][line width=0.75]    (494.5,71) -- (428.5,281) ;

\draw [color={rgb, 255:red, 155; green, 155; blue, 155 }  ,draw opacity=1 ] [dash pattern={on 4.5pt off 4.5pt}]  (201.5,424) -- (296.5,424) ;

\draw  (187.5,281.22) -- (560.5,281.22)(201.56,44) -- (201.56,436) (553.5,276.22) -- (560.5,281.22) -- (553.5,286.22) (196.56,51) -- (201.56,44) -- (206.56,51)  ;

\draw (202,23) node   {$t$};
\draw (499,168) node   {$\lambda _{0}( t)$};
\draw (578,282) node   {$X$};
\draw (227,331) node   {$\lambda _{1}( t)$};
\draw (340,201) node   {$H^{-1}( t)$};
\draw (154,276) node   {$t_{R}$};
\draw (149,415) node   {$t_{i}$};
\draw (159,74) node   {$t_{0} \sim \frac{1}{H_{0}}$};

\end{tikzpicture}

\caption{Space-time sketch of inflationary cosmology. The vertical axis is time, the horizontal axis
represents physical distance. The inflationary period lasts from $t_i$ to $t_R$. Shown are the
Hubble radius $H^{-1}(t)$ and two length scales $\lambda_0(t)$ and $\lambda_1(t)$ (fixed wavelength
in comoving coordinates). For inflation to provide a possible explanation for the observed fluctuations
on large scales, the scale $\lambda_0(t)$ corresponding to the current Hubble horizon must originate inside
of the Hubble radius at the beginning of inflation. This leads to the condition (\ref{success}). The TCC, on
the other hand, demands that the length scale $\lambda_1(t)$which equals the Hubble radius at the end of inflation was never trans-Planckian. In the sketch, both conditions are marginally satisfied.}
\end{center} 
\end{figure}

In the approximation of constant value of $H$ during inflation, the TCC condition (\ref{TCC}) can be
written in the form
\be \label{TCC2}
e^{N_+} \, < \frac{ M_{pl}}{H} \, .
\ee
The equation (\ref{success2}) for $N_+$ and the upper bound (\ref{TCC2}) on $N_+$ coming from the TCC are compatible only provided that the condition
\be
V^{3/4} \, < \sqrt{3}\, M_{pl}^2 (T_0 T_{eq})^{1/2} 
\ee
is satisfied. Inserting the values of $T_0$, $T_{eq}$ and $M_{pl}$ we obtain
\be \label{Vbound}
V^{1/4} \, < 6\times\, 10^8 {\rm{GeV}}\sim 3\times 10^{-10}M_{pl} \, .
\ee
Note that this conclusion is independent of the assumption that quantum fluctuations during inflation are
the seeds for primordial structure formation. While we have used a potential $V$ to describe the energy density during inflation, our analysis holds for more general scenarios and Eq.~(\ref{Vbound}) can be interpreted as a bound on the energy density during the inflationary epoch.

We now add the assumption that quantum fluctuations of the inflaton are responsible for the origin of structure. In this case,
the power spectrum ${\cal{P}}$ of the curvature fluctuation ${\cal{R}}$ (see \cite{MFB}) is given by
\be \label{spec}
{\cal{P}}_{\cal{R}}(k) \,  = \, \frac{1}{ 8\pi^2\epsilon} \bigl( \frac{H(k)}{M_{pl}} \bigr)^2 \, ,
\ee
where $k$ is the comoving wavenumber of the fluctuation mode and $H(k)$ is the value of $H$
at the time when the mode $k$ exits the Hubble radius. The parameter $\epsilon$ determines
the deviation of the equation of state in the inflationary phase compared to pure de Sitter:
\be
\epsilon \, \equiv \, \frac{3}{2} \bigl( \frac{p}{\rho} + 1 \bigr) \, ,
\ee
where $p$ and $\rho$ are pressure and energy densities, respectively. For inflation to provide the source of structure in the Universe, we need \cite{Planck}
\be \label{value}
{\cal{P}}_{\cal{R}}(k) \, \sim \, 10^{-9} \, .
\ee
Combining (\ref{Vbound}) (\ref{spec}) and (\ref{value}) leads to an upper bound on $\epsilon$
\begin{align}\label{epsilon}
\epsilon \, \sim \,10^9 \frac{1}{8\pi^2 }\bigl( \frac{H(k)}{M_{pl}} \bigr)^2 \, \sim \, 10^9 \frac{V}{24\pi^2M_{pl}^4} \, < \, 10^{-31} \, .
\end{align}

Since the power spectrum of gravitational waves is given by
\be \label{valued}
{\cal{P}}_{h}(k) \, \sim \,   \bigl( \frac{H(k)}{M_{pl}} \bigr)^2 \, ,
\ee
the tensor to scalar ratio $r$ is given by
\be \label{result}
r \, = \, 16 \epsilon \, < \, 10^{-30} \, ,
\ee
where the factor $16$ comes from the different normalization conventions for the scalar and tensor
spectra. While the discussion above assumed that the inflaton dominated the scalar perturbations it is important to note that the TCC constrains the absolute amplitude of the primordial gravitational waves. The bound on $r$ therefore relies only on the TCC bound on the energy in Eq.~(\ref{Vbound}) and the observed amplitude of $\mathcal{P}_{\mathcal{R}}$. Allowing scalar perturbations from additional fields or a modified sound speed for the inflaton, for example, will not relax Eq.~(\ref{result}).

From (\ref{result}) we draw the conclusion that any detection of primordial gravitational waves on
cosmological scales would provide evidence for a different origin of the primordial gravitational wave
spectrum than any inflationary model consistent with the TCC. Note that a
number of cosmological scenarios alternative to inflation do predict  significant primordial tensor
modes on cosmological scales. One example is String Gas Cosmology which predicts both a
scale-invariant spectrum of cosmological perturbations with a slight red tilt \cite{NBV} and a
roughly scale-invariant spectrum of gravitational waves with a slight blue tilt \cite{BNPV2}. 

Note that the above analysis applies not only to single field inflation, but also to multi-field
inflation. The conclusions only depend on the fact that the parameter $\epsilon$ is $\ll 1$ which is self-consistent with what we found. The constraint also
applies to {\it warm inflation} \cite{warm} models, models which can be consistent with
the de Sitter swampland conjecture \cite{Kamali}.
 
In this section, we have been general and have not assumed a slow-roll inflation.
In the following section we will study the consequences of the TCC for slow-roll inflation.

\section{Application to Slow-Roll Inflation} 

The equation of motion of a canonically normalized scalar field $\phi$ in a homogeneous and isotropic metric of the form (\ref{FRW}) is 
\be
{\ddot{\phi}} + 3 H {\dot{\phi}} + V' \, = \, 0 \, ,
\ee
where the prime indicates the derivative with respect to $\phi$. Here, we have assumed that there is no important coupling of $\phi$ to other matter fields during inflation. Thus, the analysis in this section applies to cold inflation but not to warm inflation. The Friedmann equation takes the form
\be
3 H^2 M_{pl}^2 \, = \frac{1}{2} {\dot{\phi}}^2 + V \, ,
\ee

In the case of single field slow-roll inflation, the slow-roll parameter $\epsilon$ is
\be \label{SRepsilon}
\epsilon \, \simeq \, \frac{M_{pl}^2}{2} \bigl( \frac{V^{'}}{V} \bigr)^2 \, .
\ee
The slow-roll equation of motion is
\be
3 H {\dot{\phi}} \, = \, - V^{'} \, .
\ee

The field range $\Delta \phi$ which the inflaton field $\phi$ moves during the period of
inflation is given by
\be
|\Delta \phi| \, \simeq \, |{\dot{\phi}} \Delta t| \, ,
\ee
where $\Delta t$ is the time period of inflation. We will show that $\Delta \phi$ is very small
compared to the Planck mass. In this case, it is self-consistent to assume that $H$ and ${\dot{\phi}}$
are constant. In this case using the TCC
\be
\Delta t \, = \, H^{-1} N \, < \, H^{-1} {\rm{ln}} \bigl( \frac{ M_{pl}}{H} \bigr) \, ,
\ee
and, making use of (\ref{SRepsilon}), the field range becomes
\begin{align} \label{range}
 |\Delta \phi| \, < &\sqrt{2}\epsilon^{1/2}  {\rm{ln}} \bigl( \frac{M_{pl}}{H} \bigr) M_{pl}\nonumber \\
 <&{10^{9/2}V^{1/2} \over M_{pl}}\ln( {M^2_{pl}\over {\sqrt V}})\nonumber \\
 < &10^{-13}M_{pl}\, ,
 \end{align}
where in the last inequality we used the monotonicity of $[x\ln(1/x)]$ for $ x<e$ to substitute the upper bound on $V $ from \eqref{Vbound}.
As first studied in \cite{Kung}, in the case of {\it large field inflation}, i.e. $|\Delta \phi| \gg M_{pl}$, the
inflationary slow-roll trajectory is a local attractor in initial condition space, even taking into account metric
fluctuations \cite{Hume,East}. Small field inflation as is the case here, on the other hand, is not an attractor in initial
condition space, as reviewed in \cite{Goldwirth}. If the field range for slow-roll inflation is constrained
by the TCC conjecture to obey (\ref{range}), then the inflationary scenario is faced with an initial
condition problem. The expected initial field velocity is
\be
{\dot{\phi}}_i^2 \, \sim \, V \, 
\ee
and hence
\be
\frac{{\dot{\phi}}_{SR}}{{\dot{\phi}}_{i}} \, \sim \, \epsilon^{1/2}< 10^{-15}\, ,
\ee
and it takes fine tuning of the initial velocity in order to be sufficiently close to the slow-roll trajectory. 

In the following we propose a model which can consistently explain observations, including the observational value of the tilt. We consider an inverted parabola potential $V(\phi)=V_0-{|V''|\phi^2/2}$ over a small field range $[\phi_i,\phi_f]$ such that $\delta V/V\ll1$ over the field range. Given the smallness of $\epsilon$ from \eqref{epsilon}, we have
\begin{align}\label{M1}
M_{pl}^2\frac{V''}{V}\simeq \frac{n_s-1}{2},
\end{align}
where $n_s=1+2\eta-6\epsilon$ is the tilt parameter and $\eta=M_{pl}^2V''/V$ is the second slow-roll parameter. This fixes $V''$ from observation. From the equations \eqref{spec} and \eqref{SRepsilon} we find
\begin{align}\label{M2}
\frac{V_0}{12\pi^2M_{pl}^2\phi_{CMB}^2}\simeq\mathcal{P}(k)(\frac{n_s-1}{2})^2.
\end{align}
where $\phi_{CMB}$ is the value of the field when the modes on CMB scales exited the Hubble horizon. Assuming $H$ remains almost constant $H\simeq \sqrt{V_0/3M_{pl}^2}$ during the slow-roll inflation, one can show
\begin{align}\label{e-fold2}
 \ln(\frac{\phi_2}{\phi_1})&\simeq|\eta|N(\phi_1\rightarrow \phi_2),
\end{align}
where $N(\phi_1\rightarrow \phi_2)$ is the number of e-folds accrued as $\phi$ goes from $\phi_1$ to $\phi_2$. If we plug $\phi_1=\phi_{CMB}$ and $\phi_2=\phi_{f}$ into the above identity and use the equation \eqref{success2}, we find
\begin{align}\label{efold}
    \frac{\phi_{f}}{\phi_{CMB}}=e^{|\eta|N_+}\simeq\frac{V_0^\frac{|\eta|}{4}\Omega_m^\frac{|\eta|}{2}}{(T_0T_{eq})^\frac{|\eta|}{2}}.
\end{align}
Plugging $\phi_{CMB}$ from \eqref{M2} into \eqref{efold} leads to
\begin{align}\label{end}
    \phi_{f}\simeq \frac{V_0^{\frac{1}{2}+\frac{|\eta|}{4}}\Omega_m^\frac{|\eta|}{2}}{M_{pl}^2(T_0T_{eq})^\frac{|\eta|}{2}12^\frac{1}{2}\mathcal{P}^\frac{1}{2}\pi|\eta|}\simeq 3.9 \times10^5\cdot \left(\frac{V_0}{M_{pl}}\right)^{0.505},
\end{align}
where in the last step we substituted $|\eta|\simeq0.02$, $\mathcal{P}\simeq2\cdot10^{-9}$, $T_0\simeq3K$, $T_{eq}\sim10^4K$, and $\Omega_m\simeq0.3$. This fixes the end of the field range $\phi_f$ in terms of the energy scale $V_0$. The only free parameters left are $V_0$ and $\phi_i$. 

Now we impost the TCC for the slow-roll trajectory to find a constraint in terms of $\phi_i$ and $V_0$. Plugging $\phi_1=\phi_i$ and $\phi_2=\phi_{CMB}$ in \eqref{e-fold2} gives
\begin{align}\label{fieldefold}
\phi_{CMB}\simeq \phi_i e^{\frac{1-n_s}{2}N_-},
\end{align}
where $N_{-}$ is the number of e-folds accrued before the modes on CMB scales exit the horizon. The total number of e-foldings is $N_{total}=N_{-}+N_{+}$. From \eqref{success2} we find
\begin{align}
e^N\simeq e^{N_-}\frac{V^{1/4}}{(T_0 T_{eq})^{1/2}}.
\end{align}
On the other hand, from the TCC, we know that the total number of e-folds is bounded by $e^N<M_{pl}/H$. Using \eqref{fieldefold}, this can be expressed as
\begin{align}\label{ineqfield}
(\frac{\phi_{CMB}}{\phi_i})^\frac{2}{1-n_s}<\frac{(3M_{pl}^4T_0T_{eq})^\frac{1}{2}}{V_0^\frac{3}{4}}.
\end{align}
Plugging $\phi_{CMB}$ from \eqref{M2} in \eqref{ineqfield} with $\mathcal{P}\simeq 2\cdot10^{-9}$, $n_s\simeq 0.96$, $T_0\simeq 3K$, and $T_{eq}\sim 10^4 K$ leads to
\begin{align}\label{TCCM}
(\frac{V_0}{M_{pl}^4})^{1.03}<6.6\times10^{-12}\cdot(\frac{\phi_i}{M_{pl}})^2. 
\end{align}

The above inequality is necessary for the potential to be consistent with the TCC, but it is not sufficient. This is because a potential is consistent with the TCC if the inequality \eqref{TCC} is satisfied for every expansionary trajectory, not just one particular trajectory. 

For energy scale $V_0^{1/4}=10^{-10}M_{pl}$ the potential $V_0(1-0.02\phi^2)$ defined over the field range $[\phi_i,\phi_f]=[9.7\times 10^{-16}M_{pl},2.4\times 10^{-15}M_{pl}]$ satisfies all the criteria \eqref{Vbound}, \eqref{end}, and \eqref{TCCM}. These criteria were imposed by observation and consistency with \eqref{TCC} for the slow-roll trajectory. By numerical analysis, we further verified the consistency of this potential with the inequality \eqref{TCC} for every expansionary trajectory. This is an example of a simple potential that can explain the observation and be consistent with the TCC at the same time, however, due to its short field range, it suffers from the fine-tuning problem. 

\section{Conclusions and Discussion} \label{conclusion}

We have studied the implications of the recently proposed the TCC for inflationary cosmology. Demanding that the TCC holds and that the largest scales that we currently probe in cosmology are sub-Hubble at the beginning of the inflationary phase (a necessary condition for the causal generation mechanism of fluctuations of inflationary cosmology to work) leads to an upper bound on the energy scale of inflation which is of the order of $10^9 {\rm{GeV}}$. Demanding that the amplitude of the cosmological perturbations agree with observations then leads to an upper bound on the generalized slow-roll parameter $\epsilon$ of the order of $\epsilon < 10^{-31}$. As a consequence, the tensor to scalar ratio is predicted to be smaller than $10^{-30}$. A detection of primordial gravitational waves via B-mode polarization, pulsar timing measurements or direct detection, assuming the TCC, would then imply that the source of these gravitational waves is {\it not} due to quantum fluctuations during inflation.

The above conclusions are independent of any assumptions on the possible single-field nature of inflation. If we then specialize the discussion to the case of single field slow-roll inflation with a canonically normalized inflaton field, we find that the range $\Delta \phi$ which the inflaton field traverses during the inflationary phase is of the order of $\epsilon^{1/2} M_{pl}$. This raises an initial condition problem for most of the models since the expected field velocity is much larger than the field velocity along the slow-roll trajectory. 

We proposed an inverted parabola potential as a simple example that is consistent with the TCC and can explain the observation at the same time.

\section*{Acknowledgement}

\noindent We thank the Simons Center for Geometry and Physics for hospitality during the 2019 Summer Workshop on String Theory during which the ideas for this project were developed. RB thanks the Institute for Theoretical Physics of the ETH for hospitality. The research at McGill is supported in part by funds from NSERC and from the Canada Research Chair program. ML is supported by DOE DE-SC0017848. The research of CV is supported in part by the NSF grant PHY-1719924 and by a grant from the Simons Foundation (602883, CV).


\begin{thebibliography}{99} 
  
\bibitem{Gorski:1996cf}
 K.~M.~Gorski, A.~J~Banday, C.~L.~ Bennett, G.~Hinshaw, A.~ Kogut, G.~F.~Smoot, E.~L.~ and Wright, ``Power spectrum of primordial inhomogeneity determined from the four year COBE DMR sky maps," 
Astrophys. J. 464 L11 (1996) [arXiv:astro-ph/9601063]
  
\bibitem{Bennett:2012zja}
C.~L.~Bennett, C. L. and others, ``Nine-Year Wilkinson Microwave Anisotropy Probe (WMAP) Observations: Final Maps and Results,"
Astrophys. J. Suppl. 20, 208 (2013)
[arXiv:1212.5225]

\bibitem{Aghanim:2019ame}
N. Aghanim, N. and others, ``Planck 2018 results. V. CMB power spectra and likelihoods", (2019) [arXiv:1907.12875]

\bibitem{Guth}
A.~H.~Guth,
 ``The Inflationary Universe: A Possible Solution to the Horizon and Flatness Problems,''
 Phys.\ Rev.\ D {\bf 23}, 347 (1981)
 [Adv.\ Ser.\ Astrophys.\ Cosmol.\  {\bf 3}, 139 (1987)].
 doi:10.1103/PhysRevD.23.347;\\
    A.~D.~Linde,
 ``A New Inflationary Universe Scenario: A Possible Solution of the Horizon, Flatness, Homogeneity, Isotropy and Primordial Monopole Problems,''
  Phys.\ Lett.\  {\bf 108B}, 389 (1982)
  [Adv.\ Ser.\ Astrophys.\ Cosmol.\  {\bf 3}, 149 (1987)];\\
 R.~Brout, F.~Englert and E.~Gunzig,
 ``The Creation Of The Universe As A Quantum Phenomenon,''
 Annals Phys.\  {\bf 115}, 78 (1978);\\
 A.~A.~Starobinsky,
 ``A New Type Of Isotropic Cosmological Models Without Singularity,''
 Phys.\ Lett.\ B {\bf 91}, 99 (1980);\\
 K.~Sato,
 ``First Order Phase Transition Of A Vacuum And Expansion Of The Universe,''
 Mon.\ Not.\ Roy.\ Astron.\ Soc.\  {\bf 195}, 467 (1981).
  

\bibitem{RHBrev}
R.~H.~Brandenberger,
 ``Alternatives to the inflationary paradigm of structure formation,''
 Int.\ J.\ Mod.\ Phys.\ Conf.\ Ser.\  {\bf 01}, 67 (2011)
 doi:10.1142/S2010194511000109
 [arXiv:0902.4731 [hep-th]].
 
\bibitem{ChibMukh}
V. Mukhanov and G. Chibisov,
 ``Quantum Fluctuation And Nonsingular Universe. (In Russian),''
 JETP Lett.\  {\bf 33}, 532 (1981) [Pisma Zh.\ Eksp.\ Teor.\ Fiz.\  {\bf 33}, 549 (1981)].
 
\bibitem{Starob}
A.~A.~Starobinsky,
``Spectrum of relict gravitational radiation and the early state of the universe,''
  JETP Lett.\  {\bf 30}, 682 (1979)
  [Pisma Zh.\ Eksp.\ Teor.\ Fiz.\  {\bf 30}, 719 (1979)].
  
\bibitem{Jerome}
J.~Martin and R.~H.~Brandenberger,
  ``The TransPlanckian problem of inflationary cosmology,''
  Phys.\ Rev.\ D {\bf 63}, 123501 (2001)
  doi:10.1103/PhysRevD.63.123501
  [hep-th/0005209].
  
\bibitem{Niemeyer}
J.~C.~Niemeyer,
  ``Inflation with a Planck scale frequency cutoff,''
  Phys.\ Rev.\ D {\bf 63}, 123502 (2001)
  doi:10.1103/PhysRevD.63.123502
  [astro-ph/0005533].
  
\bibitem{Greene}
R.~Easther, B.~R.~Greene, W.~H.~Kinney and G.~Shiu,
  ``Inflation as a probe of short distance physics,''
  Phys.\ Rev.\ D {\bf 64}, 103502 (2001)
  doi:10.1103/PhysRevD.64.103502
  [hep-th/0104102];\\
  A.~Kempf and J.~C.~Niemeyer,
  ``Perturbation spectrum in inflation with cutoff,''
  Phys.\ Rev.\ D {\bf 64}, 103501 (2001)
  doi:10.1103/PhysRevD.64.103501
  [astro-ph/0103225];\\
 V.~Bozza, M.~Giovannini and G.~Veneziano,
  ``Cosmological perturbations from a new physics hypersurface,''
  JCAP {\bf 0305}, 001 (2003)
  doi:10.1088/1475-7516/2003/05/001
  [hep-th/0302184].
  
\bibitem{Bedroya:2019snp} 
  A.~Bedroya and C.~Vafa,
  ``Trans-Planckian Censorship and the Swampland,''
  arXiv:1909.11063 [hep-th].

\bibitem{swamp}
H.~Ooguri and C.~Vafa,
 ``On the Geometry of the String Landscape and the Swampland,''
 Nucl.\ Phys.\ B {\bf 766}, 21 (2007)
 doi:10.1016/j.nuclphysb.2006.10.033
 [hep-th/0605264];\\
 G.~Obied, H.~Ooguri, L.~Spodyneiko and C.~Vafa,
 ``De Sitter Space and the Swampland,''
 arXiv:1806.08362 [hep-th];
 
\bibitem{swamprevs}
T.~D.~Brennan, F.~Carta and C.~Vafa,
 ``The String Landscape, the Swampland, and the Missing Corner,''
 PoS TASI {\bf 2017}, 015 (2017)
 doi:10.22323/1.305.0015
 [arXiv:1711.00864 [hep-th]];\\
 E.~Palti,
 ``The Swampland: Introduction and Review,''
 arXiv:1903.06239 [hep-th].
 
\bibitem{BV}
 R.~H.~Brandenberger and C.~Vafa,
 ``Superstrings In The Early Universe,'' Nucl.\ Phys.\ B {\bf 316}, 391 (1989).
 
\bibitem{FB}
 F.~Finelli and R.~Brandenberger,
 ``On the generation of a scale-invariant spectrum of adiabatic  fluctuations in cosmological
 models with a contracting phase,''
 Phys.\ Rev.\  D {\bf 65}, 103522 (2002) [arXiv:hep-th/0112249].
 
\bibitem{PBB}
M.~Gasperini and G.~Veneziano,
  ``Pre - big bang in string cosmology,''
  Astropart.\ Phys.\  {\bf 1}, 317 (1993)
  doi:10.1016/0927-6505(93)90017-8
  [hep-th/9211021].
  
\bibitem{Ekp}
J.~Khoury, B.~A.~Ovrut, P.~J.~Steinhardt and N.~Turok,
 ``The Ekpyrotic universe: Colliding branes and the origin of the hot big
 bang,''
 Phys.\ Rev.\ D {\bf 64}, 123522 (2001) [hep-th/0103239].

\bibitem{MFB}
 V.F. Mukhanov, H.A. Feldman and R.H. Brandenberger,
 ``Theory of Cosmological Perturbations''
 Physics Reports \textbf{215}, 203 (1992).
 
\bibitem{RHBfluctrev}
 R.~H.~Brandenberger,
 ``Lectures on the theory of cosmological perturbations,''
 Lect.\ Notes Phys.\  {\bf 646}, 127 (2004)
 doi:10.1007/978-3-540-40918-25
 [hep-th/0306071].
 
\bibitem{Planck}
N.~Aghanim {\it et al.} [Planck Collaboration],
  ``Planck 2018 results. VI. Cosmological parameters,''
  arXiv:1807.06209 [astro-ph.CO].
  
\bibitem{NBV}
 A.~Nayeri, R.~H.~Brandenberger and C.~Vafa,
  ``Producing a scale-invariant spectrum of perturbations in a Hagedorn phase of string cosmology,''
  Phys.\ Rev.\ Lett.\  {\bf 97}, 021302 (2006)
  doi:10.1103/PhysRevLett.97.021302
  [hep-th/0511140].
  
\bibitem{BNPV2}
R.~H.~Brandenberger, A.~Nayeri, S.~P.~Patil and C.~Vafa,
  ``Tensor Modes from a Primordial Hagedorn Phase of String Cosmology,''
  Phys.\ Rev.\ Lett.\  {\bf 98}, 231302 (2007)
  doi:10.1103/PhysRevLett.98.231302
  [hep-th/0604126].

\bibitem{warm}
A.~Berera,
  ``Warm inflation,''
  Phys.\ Rev.\ Lett.\  {\bf 75}, 3218 (1995)
  doi:10.1103/PhysRevLett.75.3218
  [astro-ph/9509049].
  
\bibitem{Kamali}
S.~Das,
  ``Warm Inflation in the light of Swampland Criteria,''
  Phys.\ Rev.\ D {\bf 99}, no. 6, 063514 (2019)
  doi:10.1103/PhysRevD.99.063514
  [arXiv:1810.05038 [hep-th]];\\
M.~Motaharfar, V.~Kamali and R.~O.~Ramos,
  ``Warm inflation as a way out of the swampland,''
  Phys.\ Rev.\ D {\bf 99}, no. 6, 063513 (2019)
  doi:10.1103/PhysRevD.99.063513
  [arXiv:1810.02816 [astro-ph.CO]].
  
\bibitem{Kung}
R.~H.~Brandenberger and J.~H.~Kung,
  ``Chaotic Inflation as an Attractor in Initial Condition Space,''
  Phys.\ Rev.\ D {\bf 42}, 1008 (1990).
  doi:10.1103/PhysRevD.42.1008
  
\bibitem{Hume}
H.~A.~Feldman and R.~H.~Brandenberger,
  ``Chaotic Inflation With Metric and Matter Perturbations,''
  Phys.\ Lett.\ B {\bf 227}, 359 (1989).
  doi:10.1016/0370-2693(89)90944-1
               
\bibitem{East}
W.~E.~East, M.~Kleban, A.~Linde and L.~Senatore,
  ``Beginning inflation in an inhomogeneous universe,''
  JCAP {\bf 1609}, 010 (2016)
  doi:10.1088/1475-7516/2016/09/010
  [arXiv:1511.05143 [hep-th]];\\
  K.~Clough, E.~A.~Lim, B.~S.~DiNunno, W.~Fischler, R.~Flauger and S.~Paban,
  ``Robustness of Inflation to Inhomogeneous Initial Conditions,''
  JCAP {\bf 1709}, 025 (2017)
  doi:10.1088/1475-7516/2017/09/025
  [arXiv:1608.04408 [hep-th]].
                        
\bibitem{Goldwirth}   
D.~S.~Goldwirth and T.~Piran,
  ``Initial conditions for inflation,''
  Phys.\ Rept.\  {\bf 214}, 223 (1992).
  doi:10.1016/0370-1573(92)90073-9

             
\end{thebibliography}
\end{document}